\documentclass[submission,copyright,creativecommons]{eptcs}
\providecommand{\event}{LFMTP 2026} 

\usepackage{iftex}

\ifpdf
  \usepackage{underscore}         
  \usepackage[T1]{fontenc}        
\else
  \usepackage{breakurl}           
\fi

\usepackage{hyperref}
\usepackage{amssymb,amsmath,amsthm,mathtools}
\usepackage{graphicx}
\usepackage{xcolor}
\usepackage{bussproofs}
\usepackage{url}
\usepackage{cite}
\usepackage{multicol,multirow}
\usepackage{enumitem}
\usepackage{tikz}
\usetikzlibrary{arrows.meta,positioning,shapes,automata,chains,fit,decorations.pathreplacing}
\tikzset{myarrow/.tip={_[sep=-1.4pt].To[length=2.5pt]}}  
\newcommand{\keywords}[1]{\par\addvspace\baselineskip
\noindent\textbf{Keywords:}\enspace\ignorespaces#1}

\DeclareMathAlphabet{\mathcal}{OMS}{cmsy}{m}{n}

\usepackage{fontawesome}

\newcommand{\PVSLink}{\includegraphics[scale=.15]{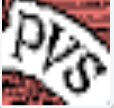}}

\newcommand{\ExtPaperLink}{\faBook}

\newcommand{\linkTHAntiunif}[4]{\href{https://github.com/nasa/pvslib/tree/0835a23b168c9e03d29e470fdd7a05e735172dcc/TRS/#1\#L#2-L#3}{{\color{blue}#4}\,\PVSLink}}

\newcommand{\linkTHAdvOnCompleteness}[4]{\href{https://github.com/mayalarincon/anti_unification_PVS/tree/main/anti_unification_LFMTP_2026/#1\#L#2-L#3}{{\color{blue}#4}\,\PVSLink}}

\newcommand{\linkExtendedPaper}[1]{\href{https://github.com/mayalarincon/anti_unification_PVS/tree/main}{{\color{blue}#1}\,\ExtPaperLink}}

\newcommand{\linkRTC}[4]{\href{https://github.com/nasa/pvslib/blob/62b35bdb113849f2ed519166e8bb778be18a695f/TRS/#1\#L#2-L#3}{{\color{blue}#4}\,\PVSLink}}

\newcommand{\unsolved}[2]{\ensuremath{\mathcal{#1}_{#2\mathtt{Uns}}}}
\newcommand{\solved}[2]{\ensuremath{\mathcal{#1}_{#2\mathtt{Sol}}}}
\newcommand{\substitution}[2]{\ensuremath{\mathcal{#1}_{#2\mathtt{Sub}}}}
\newcommand{\confdef}[2]{\ensuremath{\langle\unsolved{#1}{#2}\,\,\vert\,\,\solved{#1}{#2}\,\,\vert\,\,\substitution{#1}{#2}\rangle}}
\newcommand{\conf}[3]{\ensuremath{\langle#1\,\,\vert\,\,#2\,\,\vert\,\,#3\rangle}}

\newcommand{\vars}[1]{\ensuremath{\mathtt{vars}(#1)}}
\newcommand{\RVar}[1]{\ensuremath{\mathtt{rvars}(#1)}}

\newcommand{\Dom}[1]{\ensuremath{\mathtt{dom}(#1)}}

\newcommand{\sep}{\hspace{2mm} | \hspace{2mm}}
\newcommand{\unit}{()}

\newcommand{\SYNTname}{\ensuremath{\mathtt{Synt}}}
\newcommand{\SOLNRname}{\ensuremath{\mathtt{SolNR}}}
\newcommand{\SOLRname}{\ensuremath{\mathtt{SolR}}}
\newcommand{\DECFname}{\ensuremath{\mathtt{DecF}}}
\newcommand{\DECPname}{\ensuremath{\mathtt{DecP}}}

\newcommand{\labs}[1]{\ensuremath{\mathtt{lbls}(#1)}}

\newcommand{\auEq}[3]{\ensuremath{#2\mathrel{{\triangleq}_{{\scriptstyle#1}}} #3}}
\newcommand{\auEqIndx}[4]{\ensuremath{#2_{#4}\underset{#1_{#4}}{\triangleq} #3_{#4}}}
\newcommand{\auEqSet}[1]{\ensuremath{\{\auEqIndx{x}{s}{t}{i}\mid i\leq n\}}}

\newcommand{\lhs}[1]{\mbox{\tt lhs}(#1)}
\newcommand{\rhs}[1]{\mbox{\tt rhs}(#1)}

\newcommand{\derives}{\mathbin{\Rightarrow^{\ast}}}
\newcommand{\OneStep}[1]{\Rightarrow}
\newcommand{\SetComp}[2]{\left\{\left.\,#1\,\right\vert\,#2\,\right\}}
\newcommand{\ext}[2]{\ensuremath{\mathcal{#1}_{#2\mathtt{Ext}}}}
\newcommand{\tot}[2]{\ensuremath{\mathcal{#1}_{#2\mathtt{Tot}}}}
\newcommand{\Card}[1]{\ensuremath{\left|#1\right|}}
\newcommand{\Pair}[2]{\ensuremath{\left(#1,#2\right)}}
\newcommand{\Appl}[2]{\ensuremath{#1\,#2}}
\newcommand{\Binding}[2]{#1\mapsto#2}
\newcommand{\BasicSub}[2]{\left(\Binding{#1}{#2}\right)}
\newcommand{\eAUT}[1]{\mathord{eq}_{#1}}
\newcommand{\Set}[1]{\left\{\,#1\,\right\}}
\newcommand{\Protected}[1]{\operatorname{\mathtt{prtd}}{\left(#1\right)}}
\newcommand{\LateralL}[1]{\rho_{L}^{#1}}
\newcommand{\LateralR}[1]{\rho_{R}^{#1}}

\newcommand{\GLateralL}[1]{\tau_{L}^{#1}}
\newcommand{\GLateralR}[1]{\tau_{R}^{#1}}

\newcommand{\res}[2]{\left.#1\right\vert_{#2}}

\newcounter{mathenvironment}
\newtheorem{theorem}[mathenvironment]{Theorem}
\newtheorem{corollary}[mathenvironment]{Corollary}
\newtheorem{lemma}[mathenvironment]{Lemma}

\newtheorem{definition}[mathenvironment]{Definition}
\newtheorem{example}[mathenvironment]{Example}

\allowdisplaybreaks

\title{Anti-Unification Completeness Analysis in PVS}
\author{Mauricio Ayala-Rinc\'{o}n
\institute{Universidade de Brasília and Universidade Federal de Goi\'{a}s\\
Exact Sciences Institute and Institute of Mathematics and Statistics\\
Bras\'{i}lia D.F. and Goiânia, Brazil}
\and
Thaynara Arielly de Lima
\institute{Universidade Federal de Goi\'{a}s\\
Institute of Mathematics and Statistics\\
Goi\^{a}nia, Brazil}
\and
Maria J\'{u}lia Dias Lima
\institute{Universidade de Brasília\\
Graduate Program in Informatics\\
Bras\'{i}lia D.F., Brazil}
\and
Temur Kutsia
\institute{Johannes Kepler Universit\"{a}t\\
Research Institute for\\
Symbolic Computation\\
Linz, Austria}
\and
Marcos Mercandeli-Rodrigues
\institute{Universidade de Bras\'{i}lia\\
Graduate Program in Mathematics\\
Brasília D.F.,  Brazil}
}
\def\titlerunning{Anti-Unification Completeness Analysis in PVS}
\def\authorrunning{M. Ayala-Rinc\'{o}n et al.}
\begin{document}
\maketitle

\begin{abstract}
In syntactic anti-unification, one is concerned with finding the commonalities between terms, while (uniformly) abstracting their differences.
The original goal of anti-unification development in the seventies was to automate inductive reasoning.
Recent applications of anti-unification techniques include efficiently transforming sequential code into parallel code, detecting code clones, and preventing software failures.
Previous work addressed the elements required to verify, in the Prototype Verification System (PVS), termination and soundness of a functional algorithm based on inference rules for syntactic anti-unification. This paper dissects all aspects required to formally establish the completeness of the rule-based algorithm, highlighting the significant differences in the formalizations of anti-unification and unification.
\end{abstract}


\section{Introduction}

\subsubsection*{Motivation and Contextualization}

Generalization is the scientific principle of drawing broad conclusions from specific observations, aiming to identify patterns or laws that hold beyond the original cases studied. In logic, this idea of generalization is formalized by a concept called \emph{anti-unification}. 
It consists of finding the common structure and uniformly abstracting over differences when comparing two objects of a certain type, and it was first investigated and developed independently in the 1970s by Plotkin~\cite{Plotkin70} and Reynolds~\cite{Reynolds70}.

Given terms \(s\) and \(t\), one says that \(s\) is \emph{more general} than \(t\) (or that \(t\) is \emph{more specific} than \(s\)) and writes \(s\preceq t\) iff there exists a substitution \(\sigma\) such that \(s\sigma=t\).
The relation \(\mathord{\preceq}\) is known as \emph{instantiation preorder}~\cite{DBLP:books/el/RV01/BaaderS01}.
A \emph{generalizer} for two terms \(s\) and \(t\) is a term \(r\) such that \(r\preceq s\) and \(r\preceq t\).
The \(\mathord{\preceq}\)-minimal generalizers for \(s\) and \(t\) are called \emph{least general generalizers} for \(s\) and \(t\).
Thus, the syntactic anti-unification problem can be formally stated as:
For terms \(s\) and \(t\), construct their least general generalizers.
It is known that syntactic anti-unification is of type unitary~\cite{FDLLP:chapter/Lassez1988, DBLP:conf/ijcai/CernaK23}, i.e., any instance of the problem has a unique solution (up to renaming).
For instance, the terms \(X\), \(f(X,Y)\), \(f(g(X,Y),Z)\), and \(f(g(X,Y),X)\) are generalizers for \(f(g(c,d),c)\) and \(f(g(g(u,v),v),g(u,v))\), but only \(f(g(X,Y),X)\) is their least general generalizer.

The first procedural algorithms for solving the syntactic anti-unification problem were presented independently by Plotkin~\cite{Plotkin70} and Reynolds~\cite{Reynolds70} in the 1970s.
Their algorithm essentially reads the pair of input expressions, identifies the smallest positions where conflicts occur, and uniformly assigns fresh variables to those positions~\cite {Reynolds70}.
That allows the construction of the least general generalizer as being the join in a certain complete non-modular lattice whose order is \(\succeq\)~\cite{Reynolds70}.
Huet presented an algorithm in 1976 in~\cite{Huet76} by means of a recursive function \(\lambda\) that acts in the form \(\lambda(s,t)=f(\lambda(s_1,t_1),\ldots,\lambda(s_m,t_m))\) if \(s=f(s_1,\ldots,s_m)\) and \(t=f(t_1,\ldots,t_m)\) and \(\lambda(s,t)=\phi(s,t)\) otherwise, where \(\phi\) is a bijection between pairs of terms and variables (ensuring that \(\lambda\) solves conflicts uniformly).
A good exposition of that algorithm is presented by Lassez et al. in~\cite{FDLLP:chapter/Lassez1988}.

Rule-based procedures for anti-unification were presented by Pfenning~\cite{DBLP:conf/lics/Pfenning91} for the Calculus of Constructions, by Alpuente et al. \cite{DBLP:journals/entcs/AlpuenteEMO09} for order-sorted generalization, by the same authors for anti-unification modulo associativity and commutativity \cite{DBLP:journals/iandc/AlpuenteEEM14}, and by Baumgartner et al.~\cite{DBLP:conf/rta/BaumgartnerKLV13} for simply-typed lambda-terms in \(\eta\)-long \(\beta\)-normal forms, among others.
Alpuente et al.'s proofs of completeness were non-constructive and relied on the anti-unification type; in the syntactic case, for instance, on the fact that syntactic anti-unification is unitary. 
A more preferred proof of completeness for rule-based algorithms would be constructive, as the one presented in this paper.

Real-world applications of anti-unification are, for instance, the identification of regularities in sequential code to transform it into efficient parallel code~\cite{DBLP:journals/fgcs/BarwellBH18};
preventing failures and detecting errors in software~\cite{DBLP:conf/nsdi/MehtaB0BMAABK20};
repairing software bugs~\cite{DBLP:journals/pacmpl/BaderSPC19,DBLP:conf/sigsoft/WinterNBCHHWKWM22,DBLP:conf/sbes/SousaSGBD21};
detecting code cloning and plagiarism~\cite{DBLP:conf/ershov/BulychevKZ09,DBLP:conf/lopstr/VanhoofY19}; maintaining mathematical or software libraries~\cite{DBLP:conf/mkm/Huch25,DBLP:journals/corr/abs-2602-01107}; detecting similarities among chemical compounds to infer their carcinogenicity~\cite{Ar07}.
An open-source library of anti-unification algorithms (implemented in Java) for first- and second-order unranked terms, higher-order patterns, and nominal terms was presented in~\cite{DBLP:conf/jelia/BaumgartnerK14}.
The paper~\cite{DBLP:conf/ijcai/CernaK23} surveys the state-of-the-art in theory and applications of anti-unification. 

Surprisingly, the current interest in the practical applications of anti-unification techniques has not yet been followed by the development of formal, mechanically verified certificates in proof assistants.
That situation contrasts with the case for unification, where one aims to identify two expressions.
For instance, successful work related to the formalization of unification and matching in interactive theorem provers are the ones developed in~\cite{DBLP:conf/rta/Contejean04} for AC-matching (in Coq), in~\cite{DBLP:journals/mscs/Ayala-RinconSFS21} for nominal C-matching through unification with protected variables (also in Coq), in~\cite{DBLP:journals/entcs/Ayala-RinconFO16} for nominal unification, in~\cite{DBLP:conf/lopstr/Ayala-RinconFSN19} for nominal C-unification, in~\cite{GabrielCICM2023} for nominal AC-matching, and in~\cite{DBLP:conf/fscd/Ayala-RinconFSS22} for AC-unification (all in PVS).
Advancing this line of research to provide formal certificates for proof assistants for anti-unification techniques and to strike a balance between theoretical development and practical application is of utmost importance.
The only formalization of an anti-unification algorithm known to date is due to \cite{DBLP:conf/nfm/AyalaRinconLLMK25} for syntactic anti-unification.
In that work, an algorithm for syntactic anti-unification is specified in PVS, and its termination and soundness are mechanically verified.
Only the formal certificate for completeness remained to be constructed, and providing such a construction is the goal of the current work.
Completing that work is of utmost importance, for it will allow the extraction of certified executable code of a sound and complete algorithm for syntactic anti-unification.

This paper formally studies the additional requirements that are essential for proving the completeness of algorithms based on the standard anti-unification inference rules (see Figure \ref{fig:AURules}) that appear in \cite{DBLP:conf/nfm/AyalaRinconLLMK25}.\
The main motivation of this study is to understand in detail the greater complexity of proofs required for anti-unification compared to those for formalizations of unification algorithms (e.g., \cite{DBLP:conf/agp/Ruiz-ReinaAHM99,DBLP:journals/jar/Ruiz-ReinaMAH06,DBLP:journals/igpl/AvelarGMA14}).\
In \cite{DBLP:conf/nfm/AyalaRinconLLMK25}, for formalizing soundness, important differences were highlighted, particularly in the cases of detection of \textit{solved} and \textit{syntactically equal} problems. In anti-unification, two crucial aspects concern the treatment of such problems.
 The former corresponds to equational questions between terms headed by different (function) symbols. In contrast, the latter corresponds to trivial (i.e., syntactically equal atomic) problems in which the terms are either the same constant or the same variable.\  
Surprisingly, most of the formalization work (more than 90\%, as quantified in \cite{DBLP:conf/nfm/AyalaRinconLLMK25}) for verifying the anti-unification algorithm was devoted to these two kinds of apparently straightforward cases.  For obtaining a formal proof, the rigorous analysis of these cases requires more elaboration than the one usually devoted to the completeness of anti-unification in pen-and-paper proofs (e.g., \cite{Reynolds70,Plotkin70,Huet76,DBLP:conf/jelia/AlpuenteEEM14}).  

In unification, if a problem is what we call a \textit{solved} problem above in the context of anti-unification, one has a \textit{failure} because terms with different head function symbols can not be unified; and syntactically equal problems are trivially resolved using the identity substitution.\
In anti-unification, the situation is completely different.\
A solved equation does not mean a failure, but the detection of a difference in the structure of the terms being compared.\
That difference is recorded in the computed substitution of the current configuration.
In fact, unlike unification, anti-unification never fails. To solve the general case of anti-unification, one should proceed by further checking for other occurrences of the same difference, thereby detecting all possible regularities. 

In anti-unification, the decomposition of the problem is guided by mutually distinct labels associated with subproblems of the input problem and by a substitution that records the problem's structure. As soon as a solved subproblem is detected, it is stored separately. The essential elements for addressing solved and syntactic equations for formalizing soundness in \cite{DBLP:conf/nfm/AyalaRinconLLMK25} include preservation properties for the sets of labels of the unsolved and solved subproblems, and for the variables in the domain and the image of the computed substitution. 

\subsubsection*{Main Contribution}

\begin{itemize}

\item For the formalization of completeness, we discuss why the preservation properties developed for soundness together with the adjustments of generalizer notions (Subsection \ref{subsection:Adaptation-of-Generalizer-Notions}) are enough to address the inference rules for decomposition.

\item Although the required notions for dealing with the cases of solved and syntactic rules in the soundness proof were much more elaborate than those required for unification, these notions were not enough for formalizing anti-unification completeness. We introduce the required additional elements in notions such as arbitrary generalizers so that a series of invariance and preservation properties are guaranteed; among them, properties that assure that the steps of the algorithm do not change anti-unification problems, and that the partially computed solutions can always be refined into generalizers that are less general than any arbitrary generalizer (Subsection \ref{subsection:newrequiredDefinitions}). Based on these properties, a restricted notion of generalizer is introduced (Subsection \ref{subsection:Adaptation-of-Generalizer-Notions}).

\item  Finally, using the restricted notion of generalizer, we formalize (Subsection \ref{subsection:Completeness-Analysis-for-Solve-Syn-Rules}) the completeness theorem (stated as Theorem \ref{theorem::Strong-Completeness-Restricted-Generalizers})  and obtain as a corollary the main result that states the completeness of the algorithm regarding arbitrary generalizers (Corollary \ref{corollary::Strong-Completeness}).   

\end{itemize}

\subsubsection*{Organization}
Section \ref{sec:background} presents the required background following standard nomenclature on anti-unification, which is the one used in the formalization \cite{DBLP:conf/nfm/AyalaRinconLLMK25}. 
Section \ref{sec:analyticalproofs} is the kernel of the paper.
Subsection \ref{subsection:newrequiredDefinitions} presents the additional notions required to record preservation and invariance properties that compile the history of the algorithm's computation.
Subsection \ref{subsection:Adaptation-of-Generalizer-Notions} presents the adjustments required to address completeness; then, Subsection \ref{subsection:Completeness-Analysis-for-Solve-Syn-Rules} discusses how solved and syntactic subproblems are treated. Subsection \ref{subsection:Completeness-Analysis-for-Decomposition-Rules} discusses the analysis of decomposition rules. Finally, Section \ref{sec:conclusion} concludes and briefly discusses future work. Several points in the paper include links to the specification (\PVSLink). An \linkExtendedPaper{extended version of this paper} presents quantitative information on a direct approach for dealing with the decomposition inference rules.

\section{Background}\label{sec:background}

    The background related to terms, substitutions, and configurations is present in \cite{DBLP:conf/nfm/AyalaRinconLLMK25} and will be restated here to provide the proper vocabulary.\
    The set of \emph{terms} is generated by the standard nominal grammar \(s, t \Coloneqq c \sep X \sep \unit \sep \Pair{s}{t} \sep f\,s\) adapted from \cite{DBLP:journals/jar/Ayala-Rincon24}, where \(c\) stands for \emph{constants}, \(X\) stands for \emph{variables}, \(s\) and \(t\) stand for terms, \(\Pair{s}{t}\) stands for \emph{pairs} of terms, \(\unit\) stands for the \emph{unit},\footnote{The unit is useful for term flattening and for efficiently encoding finitary and variadic functions in the presence of pairs.\
    For instance, considering addition \(+\colon\mathbb{N}\rightarrow\mathbb{N}\), one can define \(\sum\colon\texttt{list}(\mathbb{N})\rightarrow\mathbb{N}\) recursively by \(\sum\unit\coloneqq0\) and \(\sum(m, L)\coloneqq m+\sum L\).}
    \(f\) stands for \emph{function symbols}, and \(fs\) stands for \emph{functional applications}.\
    It is specified in PVS by means of the abstract data type (ADT) \linkTHAntiunif{first_order_terms.pvs}{16}{23}{\tt first\_order\_term}, requiring types for constants, function symbols, and variables as parameters.\
    A \emph{basic substitution} is a binding \(\BasicSub{X}{t}\), where \(X\) is a variable and \(t\) is a term.\
    A \emph{substitution} is a finite list of basic substitutions (the \emph{identity substitution} is the empty list \(\iota\)).\
    Those are specified in PVS in the theory \linkTHAntiunif{first_order_substitution.pvs}{1}{537}{\tt first\_order\_substitution} using pairs and lists of these pairs.\
    The \linkTHAntiunif{first_order_substitution.pvs}{38}{68}{action of a substitution on a term} is standard.\
    The domain of a substitution \(\sigma\) is denoted by \linkTHAntiunif{first_order_substitution.pvs}{117}{117}{\(\Dom{\sigma}\)}.\
    The set of variables in its range is denoted by \(\RVar{\sigma}\).\
    The substitutions recognized by the predicate \linkTHAntiunif{first_order_substitution.pvs}{128}{134}{\tt nice?} defined in the previous theory are the most important and widely employed in the specification.

    \begin{example}[Niceness] A substitution \(\BasicSub{X_1}{t_1}\cdots\BasicSub{X_m}{t_m}\) is \emph{nice} iff \(\vars{t_i}\cap\Set{X_{i},\ldots,X_m}=\emptyset\) and \(X_i\neq X_j\) for all \(i,j=1,\ldots,m\) with \(i\neq j\).\
    For instance, \(\BasicSub{X}{\Pair{Y}{Z}}\BasicSub{W}{fX}\) is nice, while \(\BasicSub{X}{fX}\) is not.
    \end{example}

    An anti-unification problem is encoded by an \emph{anti-unification triple} (\emph{AUT}), which is an expression of the form \(\auEq{X}{s}{t}\), where \(s\) and \(t\) are terms, called its \emph{left-} and \emph{right-hand sides}, respectively, and \(X\) is a fresh variable for \(s\) and \(t\) called its \emph{label}.\
    That structure is specified in PVS by means of a record type \linkTHAntiunif{antiunif.pvs}{22}{26}{\tt AUT}, where record accessors for label and left- and right-hand sides are present.\
    The expression \(\eAUT{X}\) will be used to denote an AUT whose label is the variable \(X\).\
    Its left- and right-hand sides will be denoted by \(\lhs{\eAUT{X}}\) and \(\rhs{\eAUT{X}}\), respectively.\
    To solve an anti-unification problem, one must compare two terms and decompose their common structure.\
    That motivates an \emph{AUT classification} that allows no superposition:\
    An AUT \(\eAUT{X}\) is \emph{decomposable} iff \(\lhs{\eAUT{X}}\) and \(\rhs{\eAUT{X}}\) are either both pairs or both functional applications headed by the same function symbol;\ it is \emph{trivial} iff \(\lhs{\eAUT{X}}=\rhs{\eAUT{X}}\) is the unit, a constant or a variable; and it is \emph{solved} in any other case.\
    That AUT classification guides the design of the inference rules presented in Figure \ref{fig:AURules} and, in PVS, it is specified by means of unary predicates \linkTHAntiunif{antiunif.pvs}{40}{42}{\tt match\_DecF?}, \linkTHAntiunif{antiunif.pvs}{44}{46}{\tt match\_DecP?}, \linkTHAntiunif{antiunif.pvs}{48}{50}{\tt match\_Synt?}, and \linkTHAntiunif{antiunif.pvs}{52}{53}{\tt match\_Sol?} that checks whether or not an AUT matches the application of a rule.

    During the decomposition, the labels are the most general solutions to anti-unification (sub)problems.\
    Since those subproblems occur in different positions in the common structure of terms, it makes sense to label new subproblems with fresh labels to distinguish them.\
    The decomposition produces a finite number of AUTs that are described as a finite collection.\
    In PVS, that collection is specified as a \linkTHAntiunif{antiunif.pvs}{64}{64}{list of AUTs}.\
    To a finite collection of AUTs \(A\), one associates two sets whose members are variables that play completely distinct roles in the anti-unification algorithm.\
    The first one is its \emph{set of labels} \linkTHAntiunif{antiunif.pvs}{78}{81}{\(\labs{A}\)}, which collects all the labels of the members of \(A\).\
    The second one is its \emph{set of variables} \linkTHAntiunif{antiunif.pvs}{66}{69}{\(\vars{A}\)}, which collects all the variables that occur in the left- and right-hand sides of its members.\
    To refer to a collection of subproblems obtained from a problem consistently, it is necessary to define a \emph{valid set of AUTs} as being a finite set \(A\) of AUTs such that \(\labs{A}=\Card{A}\) and \(\vars{A}\cap\labs{A}=\emptyset\).\
    In PVS, valid sets of AUTs were specified by means of valid lists of AUTs, which are recognized by the unary predicate \linkTHAntiunif{antiunif.pvs}{95}{96}{\tt valid\_AUTs?}.

    As decomposition proceeds, it may be necessary to address an anti-unification problem that was previously handled but at a distinct position in the common structure.\
    Such problems are essentially the same but encoded by AUTs that differ only in their labels.\
    Those are called \emph{repeated AUTs}.\
    That can be extended to sets of AUTs:\ An AUT \(\eAUT{X}\) is \emph{repeated in} a set \(A\) of AUTs iff there exists an AUT \(\eAUT{Y}\) in \(A\) such that \(\eAUT{X}\) and \(\eAUT{Y}\) are repeated.\
    The binary predicates \linkTHAntiunif{antiunif.pvs}{139}{140}{\tt repeated\_AUT?} and \linkTHAntiunif{antiunif.pvs}{142}{145}{\tt AUT\_repeated\_in?} are specified to recognize such repetitions.

    \begin{figure}[htbp]
        \begin{prooftree}
          \AxiomC{\(\conf{\auEq{X}{f\,s}{f\,t},\, U}{S}{\sigma}\)}
           \LeftLabel{Decompose-Function  (\DECFname)} 
           \UnaryInfC{\(\conf{\auEq{\color{brown}Y}{s}{t},\, U}{S}{\sigma\,\BasicSub{X}{f{\color{brown}Y}}}\)}
        \end{prooftree}
        \begin{prooftree}
          \AxiomC{\(\conf{\auEq{X}{\Pair{s}{u}}{\Pair{t}{v}},\, U }{ S}{ \sigma }\)}
           \LeftLabel{Decompose-Pair (\DECPname)} 
           \UnaryInfC{\(\conf{ \auEq{\color{brown}Y}{s}{t},\,\auEq{\color{brown}Z}{u}{v},\, U }{ S }{ \sigma\,\BasicSub{X}{\Pair{{\color{brown}Y}}{{\color{brown}Z}}} }\)}
        \end{prooftree}
        \begin{prooftree}
          \AxiomC{\(\conf{\auEq{X}{s}{t},\, U}{ S}{ \sigma }\)}
           \LeftLabel{Solve-Repeated (\SOLRname)}
           \RightLabel{{if \(\auEq{X}{s}{t}\) is solved and \(\auEq{X^{\prime}}{s}{t}\in S\)}}
           \UnaryInfC{ \(\conf{ U}{ S}{ \sigma\,\BasicSub{X}{X^{\prime}} }\)}
        \end{prooftree}
        \begin{prooftree}
          \AxiomC{ \(\conf{\auEq{X}{s}{t},\, U}{ S}{ \sigma }\)}
           \LeftLabel{Solve-Non-Repeated (\SOLNRname)}
           \RightLabel{{if \(\auEq{X}{s}{t}\) is solved and not repeated in \(S\)}}
           \UnaryInfC{ \(\conf{ U}{ \auEq{X}{s}{t},\, S}{ \sigma }\)}
        \end{prooftree}
        \begin{prooftree}
          \AxiomC{\(\conf{\auEq{X}{s}{s},\, U}{ S}{ \sigma }\)}
           \LeftLabel{Syntactic (\SYNTname)}
           \RightLabel{{if \(\auEq{X}{s}{s}\) is trivial}}
           \UnaryInfC{\(\conf{ U}{ S}{ \sigma\,\BasicSub{X}{s} }\)}
        \end{prooftree}
        \caption{Standard anti-unification inference rules. } \label{fig:AURules}
    \end{figure}

    During the decomposition of the common term structure, one must understand the current state and the states visited to reach it.\
    In that way, the problems already computed and those remaining to be computed will be identified.\
    That is encoded by means of a \emph{valid configuration}, which is an expression of the form \(\confdef{C}{}\), where \(\substitution{C}{}\) is a substitution, \(\unsolved{C}{}\) and \(\solved{C}{}\) are valid sets of AUTs called its \emph{computed substitution} and \emph{unsolved} and \emph{solved} parts, respectively, and each of those three components satisfy the following constraints:
    \begin{enumerate}[label=\textbf{\arabic*.}]
        
        \item The sets \(\unsolved{C}{}\) and \(\solved{C}{}\) are disjoint and \(\unsolved{C}{}\cup\solved{C}{}\) is a valid set of AUTs;

        \item The set \(\solved{C}{}\) contains only solved AUTs that are not repeated in \(\solved{C}{}\);

        \item The sets \(\labs{\unsolved{C}{}}\), \(\labs{\solved{C}{}}\), and \(\Dom{\substitution{C}{}}\) are pairwise disjoint.\
        The sets \(\Dom{\substitution{C}{}}\) and \(\RVar{\substitution{C}{}}\) are also disjoint.
        
    \end{enumerate}

The inference rules in Figure \ref{fig:AURules} induce a reduction relation over valid configurations, which is denoted as $\Rightarrow$ using standard rewriting notation \cite{DBLP:books/TermRewritingAllThatBaaderNipkow}; thus, $\Rightarrow^*$, $\Rightarrow^+$, and $\Rightarrow^n$ denote the reflexive transitive closure, transitive closure, and the $n$-reduction steps derivation relations obtained from $\Rightarrow$, respectively. In specific derivations, superscripts with the names of the applied rules are added to the relation symbol $\Rightarrow$. For the example in the introduction, we have the following derivation:

\[\begin{array}{rl}
\conf{\auEq{X}
{\Appl{f}{\Pair{\Appl{g}{\Pair{c}{d}}}{c}}}
{\Appl{f}{\Pair{\Appl{g}{\Pair{\Appl{g}{\Pair{u}{v}}}{v}}}{\Appl{g}{\Pair{u}{v}}}}}}
{\emptyset}{\iota} 
& \xRightarrow{\DECFname,\,\DECPname} \\
\conf{
\auEq{Y}{\Appl{g}{\Pair{c}{d}}}{\Appl{g}{\Pair{\Appl{g}{\Pair{u}{v}}}{v}}},
\auEq{Z}{c}{\Appl{g}{\Pair{u}{v}}}}
{\emptyset}{X\mapsto \Appl{f}{\Pair{Y}{Z}}} 
& \xRightarrow{\DECFname,\,\DECPname} \\
\conf{
\auEq{Y_1}{c}{\Appl{g}{\Pair{u}{v}}},
\auEq{Y_2}{d}{v}, 
\auEq{Z}{c}{\Appl{g}{\Pair{u}{v}}}
}
{\emptyset}
{X\mapsto \Appl{f}{\Pair{\Appl{g}{\Pair{Y_1}{Y_2}}}{Z}}}
& \xRightarrow{\SOLNRname,\,\SOLNRname} \\
\conf{
\auEq{Z}{c}{\Appl{g}{\Pair{u}{v}}}
}
{\auEq{Y_1}{c}{\Appl{g}{\Pair{u}{v}}},
\auEq{Y_2}{d}{v} 
}
{X\mapsto \Appl{f}{\Pair{\Appl{g}{\Pair{Y_1}{Y_2}}}{Z}}}
& \xRightarrow{\SOLRname} \\
\conf{\emptyset
}
{\auEq{Y_1}{c}{\Appl{g}{\Pair{u}{v}}},
\auEq{Y_2}{d}{v} 
}
{X\mapsto \Appl{f}{\Pair{\Appl{g}{\Pair{Y_1}{Y_2}}}{Y_1}}}
\end{array}
\]
    
    A configuration is specified in PVS by means of \linkTHAntiunif{antiunif.pvs}{121}{121}{\tt Configuration}, which is a record type with record accessors for unsolved, solved, and computed substitution (always a nice one).\
    Among configurations, the valid ones are recognized by the unary predicate \linkTHAntiunif{antiunif.pvs}{182}{185}{\tt validConfiguration?}.\
    The intuition for each component of a valid configuration is as follows.\
    Its unsolved part encodes the problems to be computed, its solved part encodes the problems where the common term structure differs and which were detected for the first time, and its computed substitution encodes the common term structure already decomposed.\
    Thus, a valid configuration encodes the entire history of the decomposition up to that point.

    The decomposition of the term structure is carried out by the inference rules presented in Figure \ref{fig:AURules}.\
    The algorithm \linkTHAntiunif{antiunif.pvs}{454}{471}{\tt Antiunify} is specified in \cite{DBLP:conf/nfm/AyalaRinconLLMK25} as a recursive unary function that maps valid configurations into valid configurations, and it consists of the exhaustive application of the rules to a valid input configuration.\
    Since the unsolved part of a valid configuration is specified by a list of AUTs, the rules are always applied to the first member of that list.\ 
    The size of the unsolved part of a configuration, which is the sum of the \linkTHAntiunif{antiunif.pvs}{37}{38}{\tt sizes} of its members, was the chosen metric for ensuring \emph{termination} in \cite{DBLP:conf/nfm/AyalaRinconLLMK25}.

    The rules presented in Figure \ref{fig:AURules} are specified in \cite{DBLP:conf/nfm/AyalaRinconLLMK25} by means of unary functions \linkTHAntiunif{antiunif.pvs}{330}{341}{\tt  DecF}, \linkTHAntiunif{antiunif.pvs}{377}{388}{\tt DecP}, \linkTHAntiunif{antiunif.pvs}{402}{408}{\tt Synt}, and \linkTHAntiunif{antiunif.pvs}{440}{447}{\tt Solve} that require an input valid configuration that matches the application of the rule and returns an output configuration according to their (dependent) type.\
    That type encodes, among other properties, that the output configuration is a valid configuration with a \textit{smaller size} than the input configuration.\
    PVS generated a \emph{type correctness condition} (TCC) for each function, and each one of those TCCs required a manual proof.\
    The advantage of this approach is that PVS generates a termination TCC for the specification of \texttt{Antiunify}, and the prover can automatically prove it using the information encoded in the (dependent) types of the input-output configurations of the inference rules.\
    Thus, an exhaustive application of the rules terminates in a configuration with an empty unsolved part.\
    Those are called \emph{normal configurations} (or \emph{final configurations}) and are recognized in PVS by means of the predicate \linkTHAntiunif{antiunif.pvs}{525}{525}{\tt normal\_configuration?}.

    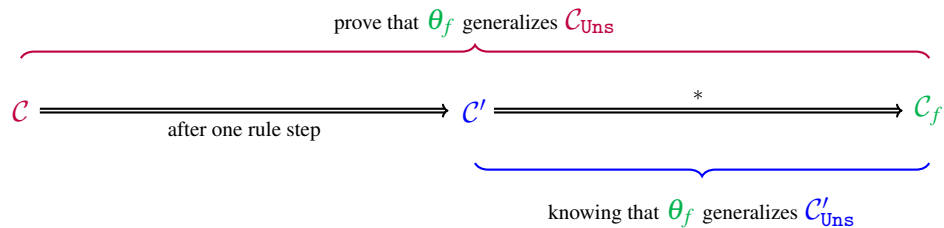
\begin{figure} [htpb]
        \centering
        \begin{tikzpicture}[scale=1]
            \draw (0,0) node {\(\boldsymbol{\color{purple}\mathcal{C}}\)};
            \draw (6,0) node {\(\boldsymbol{\color{blue}\mathcal{C}'}\)};
            \draw (12,0) node {\(\boldsymbol{\color{blue!30!green!100!black!100!}\mathcal{C}_f}\)};
            \draw[thick,double,-myarrow] (0.25,0) -- (5.65,0) node[midway, below] {\scriptsize{after one rule step}};
            \draw[thick,double,-myarrow] (6.25,0) -- (11.65,0) node[midway, above] {\scriptsize{\(\boldsymbol\ast\)}};
            \draw [color=purple,thick,decorate,decoration={brace,amplitude=5pt,raise=4ex}]
  (0,0) -- (12,0) node[midway,yshift=3em]{{\scriptsize\color{black}prove that}  \(\boldsymbol{\color{blue!30!green!100!black!100!}\theta_f}\) {\scriptsize\color{black}generalizes} \(\color{purple}\unsolved{C}{}\)};
            \draw [color=blue,thick,decorate,decoration={brace,amplitude=5pt,mirror,raise=4ex}]
  (6,0) -- (12,0) node[midway,yshift=-3.5em]{{\scriptsize\color{black}knowing that} \(\boldsymbol{\color{blue!30!green!100!black!100!}\theta_f}\) {\scriptsize\color{black}generalizes} \(\boldsymbol{\color{blue}\unsolved{C}{}'}\)};
        \end{tikzpicture}
        \caption{Inductive step in the soundness theorem presented in \cite{DBLP:conf/nfm/AyalaRinconLLMK25}.} \label{figure::Soundness}
    \end{figure}

    The background was enough to prove \linkTHAntiunif{antiunif.pvs}{599}{605}{\tt antiunif\_is\_sound} in \cite{DBLP:conf/nfm/AyalaRinconLLMK25}.\
    The argument was constructed by induction on the size of the unsolved part of a valid configuration together with a case analysis concerning the employed rule in the inductive step, which is presented in Figure \ref{figure::Soundness}.\
    The effort required was measured in \cite{DBLP:conf/nfm/AyalaRinconLLMK25} by the number of proof commands (for the proof and also its dependencies) and the amounts are \(2.09\)\% for \DECFname, \(4.56\)\% for \DECPname, \(61.70\)\% for \SYNTname, \(29.60\)\% for \SOLRname, and \(2.05\)\% for \SOLNRname.\
    Surprisingly, the rules \SYNTname\ and \SOLRname\ together make up \(91.10\)\% of the entire effort and algorithmically, going from \(\mathcal{C}\) to \(\mathcal{C}'\), they only add to \(\substitution{C}{}\) a new basic substitution \(\BasicSub{X}{a}\), where \(X\) is a label in \(\unsolved{C}{}\) and \(a\) is an atomic term.\
    While that transformation seems trivial in a pen-and-paper proof, a considerable amount of work is required to analyze its effects in a mechanized and formally verified proof for the case where \(a\) is a variable.\
    In that case, the branches for those two rules relied on dependencies (lemmas) that stated the invariance and preservation properties concerning the action of \(\theta_f\) on the variable \(a\). These properties explained why all that effort was required.

\section{Dissection Towards a Mechanized Completeness Proof}\label{sec:analyticalproofs}

Using the formalization elements developed for the proof of soundness in \cite{DBLP:conf/nfm/AyalaRinconLLMK25}, one can try to address the completeness proofs for the decomposition rules (\DECFname\ and \DECPname).\
Indeed, for an inductive proof like the one presented in Figure \ref{figure::Soundness} using a notion  \linkTHAdvOnCompleteness{completeness.pvs}{34}{41}{\tt r\_generalizer} of arbitrary generalizer that restrict its domain and range avoiding occurrences of the variables used by the algorithm, the completeness case analysis of these two rules will amount more than 2500 and 3500 PVS proof commands for the \DECFname\ and \DECPname\ rules, respectively (see the \linkExtendedPaper{extended version of this paper}).\
Such an approach is sufficient to guarantee the completeness of the algorithm for the subclass of anti-unification problems that do not require applications of the rules \SOLRname\ and \SYNTname, but not for arbitrary problems.
An important class of such problems comprises those that do not repeat variables or constants.
For those problems, the decomposition rules will only lead to solved AUTs that cannot appear more than once.
Since no variable nor constant appears repeatedly, no application of the rules \SOLRname\ or \SYNTname\ is possible.

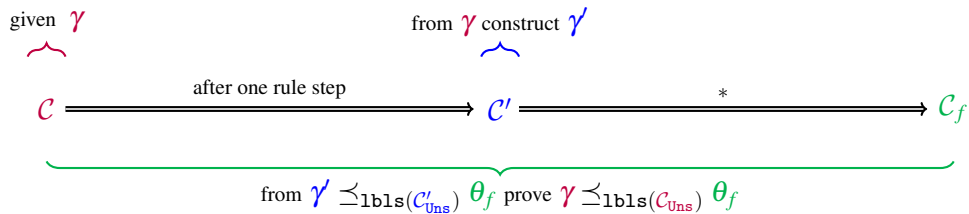
\begin{figure} [htpb]
            \centering
            \begin{tikzpicture}[scale=1]
                \draw (0,0) node {\(\boldsymbol{\color{purple}\mathcal{C}}\)};
                \draw (6,0) node {\(\boldsymbol{\color{blue}\mathcal{C}'}\)};
                \draw (12,0) node {\(\boldsymbol{\color{blue!30!green!100!black!100!}\mathcal{C}_f}\)};
                \draw[thick,double,-myarrow] (0.25,0) -- (5.65,0) node[midway, above] {\scriptsize{after one rule step}};
                \draw[thick,double,-myarrow] (6.25,0) -- (11.65,0) node[midway, above] {\scriptsize{\(\boldsymbol\ast\)}};
                \draw [color=purple,thick,decorate,decoration={brace,amplitude=5pt,raise=4ex}]
      (-.25,0) -- (.25,0) node[midway,yshift=3em]{{\scriptsize\color{black}given } 
      \({\color{purple}\boldsymbol\gamma}\) };
                \draw [color=blue,thick,decorate,decoration={brace,amplitude=5pt,raise=4ex}]
      (5.75,0) -- (6.25,0) node[midway,yshift=3em]{{\scriptsize\color{black}from} \(\boldsymbol{\color{purple}\gamma}\) {\scriptsize\color{black}construct} \(\boldsymbol{\color{blue}\gamma'}\)};
                \draw [color=blue!30!green!100!black!100!,thick,decorate,decoration={brace,amplitude=5pt,mirror,raise=4ex}]
      (0,0) -- (12,0) node[midway,yshift=-3em]{{\scriptsize\color{black}from} \(\boldsymbol{\color{blue}\gamma'\color{black}\preceq_{\labs{{\color{blue}\unsolved{C}{}'}}}\color{blue!30!green!100!black!100!}\theta_f}\) {\scriptsize\color{black}prove} \(\boldsymbol{\color{purple}\gamma\color{black}\preceq_{\labs{{\color{purple}\unsolved{C}{}}}}\color{blue!30!green!100!black!100!}\theta_f}\)};
            \end{tikzpicture}
            \caption{Inductive step in the experiments towards a completeness proof.} \label{figure::Completeness-Old}
        \end{figure}

Surprisingly, following the same inductive schema with the same notion of generalizer did not work for the \SOLRname, \SOLNRname, and \SYNTname\ rules, as this section explains. The problem arises in the inductive step: given an arbitrary generalizer $\gamma$ for the problem encoded by a configuration $\cal C$, to apply the inductive hypothesis, one needs to build an associated generalizer $\gamma'$ for the configuration ${\cal C}'$; and using the hypothesis that it is more general than the computed solution given by $\theta_f$, in the final configuration $\mathcal{C}_f$, prove that $\gamma$ is also more general than $\theta_f$ (see Figure \ref{figure::Completeness-Old}).  

Initially, Subsection \ref{subsection:newrequiredDefinitions} explains new notions that are required to compile configuration preservation properties for the analysis of the \SYNTname, \SOLRname, and \SOLNRname\ rules. These notions are used to record invariance properties throughout the history of computations. Afterward, Subsection \ref{subsection:Adaptation-of-Generalizer-Notions} presents crucial adjustments required to address completeness for those rules; finally, Subsection \ref{subsection:Completeness-Analysis-for-Solve-Syn-Rules} discusses how these adjusted notions are used in the formalization of the inductive proof.\
Although the branches of the decomposition rules can be derived from the elements developed in \cite{DBLP:conf/nfm/AyalaRinconLLMK25}, for insertion into the case analysis of this new inductive schema, they require straightforward adaptations, as explained in Subsection~\ref{subsection:Completeness-Analysis-for-Decomposition-Rules}.

    \subsection{New Notions Required for the Completeness Analysis}\label{subsection:newrequiredDefinitions}

    An \emph{initial configuration} is a valid configuration with an empty solved part and identity computed substitution, and it is recognized in PVS by means of the unary predicate \linkTHAdvOnCompleteness{antiunify_derivation.pvs}{72}{73}{\tt initial\_configuration?}.\
    In PVS, a single application of any of the rules is recognized by the binary predicate \linkTHAdvOnCompleteness{antiunify_derivation.pvs}{336}{340}{\tt Antiunify?} specified from the binary predicates \linkTHAdvOnCompleteness{antiunify_derivation.pvs}{319}{321}{\tt DecF?}, \linkTHAdvOnCompleteness{antiunify_derivation.pvs}{323}{325}{\tt DecP?}, \linkTHAdvOnCompleteness{antiunify_derivation.pvs}{327}{329}{\tt Synt?}, and \linkTHAdvOnCompleteness{antiunify_derivation.pvs}{331}{333}{\tt Solve?}. These predicates check whether a configuration is derived from another configuration by applying the respective inference rule.\
    The binary predicate {\tt RTC(Antiunify?)}, constructed using the reflexive-transitive closure of abstract reduction relations, \linkRTC{relations_closure.pvs}{109}{109}{\tt RTC(R)}, from the NASAlib theory of rewriting (see, e.g., \cite{DBLP:journals/jar/GaldinoA10,DBLP:journals/jar/OliveiraGA17}).\
    For brevity, in this section, we will also use standard rewriting notation for the specified predicate {\tt Antiunify?}; thus, \(\mathord{\OneStep{}}\) and \(\derives\) will denote {\tt Antiunify?} and {\tt RTC(Antiunify?)}, respectively.
    
    All information pertaining to the derivation of a valid configuration from an initial configuration is stored in the former.\
    All labels that were used during its derivation encode positions in the common term structure where subproblems were detected.\
    Labels are transformed at each step taken:\ New labels are generated, labels are moved from the unsolved part to the solved part, to the domain of the computed substitution, or to its range.\
    That motivates the following definition:

    \begin{definition}[Labels of a Valid Configuration]
        The \linkTHAdvOnCompleteness{antiunify_derivation.pvs}{30}{31}{\emph{set of labels}} of a valid configuration \(\mathcal{C}\) is defined as \(\labs{\mathcal{C}}\coloneqq\labs{\unsolved{C}{}}\cup\labs{\solved{C}{}}\cup\Dom{\substitution{C}{}}\).
    \end{definition}

    The information encoded by labels is essential if one aims for completeness, as labels encode positions.\
    One important preservation property is stated in the next lemma and proved by induction on the number of steps in a derivation.\
    Intuitively, all information encoded by the labels of a valid configuration is inherited by every valid configuration derived from it.\
    In particular, if \(\mathcal{C}\OneStep{}\mathcal{C}'\) by means of \SOLRname, \SOLNRname, or \SYNTname, then \(\labs{\mathcal{C}'}=\labs{\mathcal{C}}\).

    \begin{lemma}[\linkTHAdvOnCompleteness{antiunify_derivation.pvs}{386}{388}{Preservation of Labels}]\label{lemma::Preservation-of-Labels}
        If \(\mathcal{C}\derives\mathcal{C}'\), then \(\labs{\mathcal{C}}\subseteq\labs{\mathcal{C}'}\).
    \end{lemma}

    Another important class of variables in a valid configuration derived from an initial configuration are those which appear on the left- and right-hand sides of the AUTs of the latter.\
    In the decomposition of the common term structure, one can detect a trivial AUT whose both sides are the same variable, and the rule \SYNTname\ transforms that AUT accordingly, storing the following information at the computed substitution:\ The label that encodes the position where the problem was encountered must be mapped to that variable that occurs at that position.\
    That is the only possible transformation regarding those types of variables, and it motivates the following definition:

    \begin{definition}[Protected Variables of a Valid Configuration]
        The \linkTHAdvOnCompleteness{antiunify_derivation.pvs}{253}{254}{\emph{set of protected variables}} of a valid configuration \(\mathcal{C}\) is defined as \(\Protected{\mathcal{C}}\coloneqq\vars{\unsolved{C}{}\cup\solved{C}{}}\cup\left(\RVar{\substitution{C}{}}\setminus\labs{\unsolved{C}{}\cup\solved{C}{}}\right)\).
    \end{definition}

    A variable that occurs in the left- or right-hand side of an AUT of \(\mathcal{C}_0\) can occur in a subsequent derived configuration \(\mathcal{C}\) in \(\vars{\unsolved{C}{}\cup\solved{C}{}}\cup\RVar{\substitution{C}{}}\).\
    How they appear in those sets is understood by the action of the inference rules employed in the derivation.\
    One important invariance property is presented in the following lemma, which is proved by induction on the number of steps of a derivation.\
    A crucial step in that proof is understanding how the set of variables of the computed substitution of a valid configuration changes after applications of the standard inference rules.

    \begin{lemma}[\linkTHAdvOnCompleteness{antiunify_derivation.pvs}{395}{397}{Invariance of Protected Variables}]\label{lemma::Invariance-of-Protected-Variables}
        If \(\mathcal{C}\derives\mathcal{C}'\), then \(\Protected{\mathcal{C}}=\Protected{\mathcal{C}'}\).
    \end{lemma}

    An important consequence of the previous lemma is the following: the set of protected variables of any configuration derived from an initial configuration is exactly the set of variables of the unsolved part of the latter.\
    Also, it is important to notice that \(\labs{\mathcal{C}}\cap\Protected{\mathcal{C}}=\emptyset\) for any valid configuration \(\mathcal{C}\).\ 
    Another type of invariant present throughout the computation is the problem being solved.\
    To discuss that, one requires the proper vocabulary:

    \begin{definition}[Lateral Substitutions]\label{defi::lateral-substitutions}\
        \begin{enumerate}[label=\textbf{\arabic*.}]
        
            \item The \emph{left} (resp.\ \emph{right}) \emph{substitution associated with} a valid set \(A\) of AUTs is the substitution \(\LateralL{A}\) (resp.\ \(\LateralR{A}\)) with \(\Dom{\LateralL{A}}=\labs{A}\) (resp.\ \(\Dom{\LateralR{A}}=\labs{A}\)) such that \(X\LateralL{A}=\lhs{\eAUT{X}}\) (resp.\ \(X\LateralR{A}=\rhs{\eAUT{X}}\)) for every \({X}\) in \(\labs{{A}}\);

            \item The \linkTHAdvOnCompleteness{antiunify_derivation.pvs}{87}{88}{\emph{left}} and \linkTHAdvOnCompleteness{antiunify_derivation.pvs}{90}{91}{\emph{right substitutions associated with}} a valid configuration \(\mathcal{C}\) are:
            \begin{align*}
                \LateralL{\mathcal{C}}&\coloneqq\LateralL{\unsolved{C}{}\cup\solved{C}{}}\quad\quad\text{and}\quad\quad
                \LateralR{\mathcal{C}}\coloneqq\LateralR{\unsolved{C}{}\cup\solved{C}{}}.
            \end{align*}
            
        \end{enumerate}

    \end{definition}

    By definition, one can see that \(\RVar{\LateralL{\mathcal{C}}}\cup\RVar{\LateralR{\mathcal{C}}}\subseteq\Protected{\mathcal{C}}\), thus the lateral substitutions associated with \(\mathcal{C}\) are idempotent.\
    Also, their restriction to any part of \(\mathcal{C}\) is the corresponding lateral substitution associated with that part.\
    Lateral substitutions play an important role in reconstructing AUTs that encode anti-unification problems.\
    The invariance result hinted at before can be stated as the next lemma.\
    It is proved by induction on the length of the derivation, paying attention to how each inference rule \linkTHAdvOnCompleteness{antiunify_derivation.pvs}{207}{245}{modifies the lateral} and computed substitutions associated with a configuration, as well as to the role of the label of the AUT transformed by the inference rule.\

    \begin{lemma}[\linkTHAdvOnCompleteness{antiunify_derivation.pvs}{644}{649}{Problem Invariance}]\label{lemma::Problem-Invariance}
        If \(\mathcal{C}\derives\mathcal{C}'\), then \(X\substitution{C}{}'\LateralL{\mathcal{C}'}=X\substitution{C}{}\LateralL{\mathcal{C}}\) and \(X\substitution{C}{}'\LateralR{\mathcal{C}'}=X\substitution{C}{}\LateralR{\mathcal{C}}\) for every \(X\) in \(\Dom{\substitution{C}{}}\).
    \end{lemma}

    Since all information is preserved in a derivation, one can think intuitively about Lemma \ref{lemma::Problem-Invariance} by employing a jigsaw puzzle:\
    One breaks the entire picture (an AUT of the unsolved part of a valid configuration) into small pieces (AUTs obtained from that AUT by applications of inference rules to configurations) in a way that, if all of them are put together (the compositions of the computed and lateral substitutions), then one obtains the original picture again.\
    That is the \emph{problem invariance}:\ No matter how one breaks a problem into subproblems, the problem remains the same throughout the computation.\
    
    All the preservation properties presented so far can be used to completely describe the history of computation in the derivation of a valid configuration from an initial one.\
    For instance, an argument by induction is sufficient for proving the following lemma:

    \begin{lemma}\label{lemma::History-Solved}[\linkTHAdvOnCompleteness{antiunify_derivation.pvs}{656}{660}{History of the Solved Part}]\label{lemma::History-Solved-Part}
        Let \(\mathcal{C}_0\) be an initial configuration and \(\mathcal{C}\) be a valid configuration such that \(\mathcal{C}_0\derives\mathcal{C}\).\
        For every AUT \(\eAUT{X}\) in \(\solved{C}{}\), there exist two valid configurations \(\mathcal{C}'\) and \(\mathcal{C}''\) such that \(\eAUT{X}\) belongs to \(\unsolved{C}{}'\cap\solved{C}{}''\setminus\solved{C}{}'\) and \(\mathcal{C}_0\derives\mathcal{C}'\OneStep{}\mathcal{C}''\derives\mathcal{C}\), where the step \(\mathcal{C}'\OneStep{}\mathcal{C}''\) uses \SOLNRname. 
    \end{lemma}

    Another important description of the history of computation during a derivation concerns the domain of the computed substitution.\
    That history is presented in the next lemma, which can be proved by induction using Lemma \ref{lemma::Problem-Invariance}:

    \begin{lemma}\label{lemma::History-Domain}[\linkTHAdvOnCompleteness{antiunify_derivation.pvs}{663}{710}{History of the Computed Substitution}]\label{lemma::History-Computed-Substitution}
        Let \(\mathcal{C}_0\) be an initial configuration and \(\mathcal{C}\) be a configuration such that \(\mathcal{C}_0\derives\mathcal{C}\).\
        For every \(X\) in \(\Dom{\substitution{C}{}}\), exactly one of the following holds:
        \begin{enumerate}[label=\textbf{\arabic*.}]
            
            \item \(X\substitution{C}{}\) is a variable and exactly one of the following holds:

            \begin{enumerate}[label=\textbf{1.\arabic*.}]
                
                \item \(X\substitution{C}{}\) is the label of the AUT
                \[
                    \auEq{X\substitution{C}{}}{X\substitution{C}{}\LateralL{\mathcal{C}}}{X\substitution{C}{}\LateralR{\mathcal{C}}}
                \]
                that belongs to~\(\solved{C}{}\);

                \item \(X\substitution{C}{}\) is a member of \(\Protected{\mathcal{C}}\) and the AUT
                \[
                    \auEq{X}{X\substitution{C}{}\LateralL{\mathcal{C}}}{X\substitution{C}{}\LateralR{\mathcal{C}}}
                \]
                is the trivial AUT
                \[
                    \auEq{X}{X\substitution{C}{}}{X\substitution{C}{}}
                \]
                that belongs to \(\unsolved{C}{}'\) for some valid configuration \(\mathcal{C}'\) such that~\(\mathcal{C}_0\derives\mathcal{C}'\derives\mathcal{C}\);
                
            \end{enumerate}

            \item \(X\substitution{C}{}\) is either the unit or a constant and the AUT
                \[
                    \auEq{X}{X\substitution{C}{}\LateralL{\mathcal{C}}}{X\substitution{C}{}\LateralR{\mathcal{C}}}
                \]
                is the trivial AUT
                \[
                    \auEq{X}{X\substitution{C}{}}{X\substitution{C}{}}
                \]
                that belongs to \(\unsolved{C}{}'\) for some valid configuration \(\mathcal{C}'\) such that~\(\mathcal{C}_0\derives\mathcal{C}'\derives\mathcal{C}\);

            \item \(X\substitution{C}{}\) is either a functional application or a pair, and the AUT
            \[
                \auEq{X}{X\substitution{C}{}\LateralL{\mathcal{C}}}{X\substitution{C}{}\LateralR{\mathcal{C}}}
            \]
            is decomposable and belongs to \(\unsolved{C}{}'\) for some valid configuration \(\mathcal{C}'\) such that~\(\mathcal{C}_0\derives\mathcal{C}'\derives\mathcal{C}\).
            
        \end{enumerate}
    \end{lemma}

    \subsection{Adjustment of Generalizer Notions}\label{subsection:Adaptation-of-Generalizer-Notions}

        At this point, one has sufficient grounds to provide the building blocks of a definition that encodes what a generalizer for a valid configuration is.\
        Lemma \ref{lemma::Problem-Invariance} guarantees the consistency of the following definition:
    
        \begin{definition}[Extended Set of AUTs]
            The \emph{extended set of AUTs} of a valid configuration \(\mathcal{C}\) is defined as:
            \[
                \ext{C}{}\coloneqq\SetComp{\auEq{X}{X\substitution{C}{}\LateralL{\mathcal{C}}}{X\substitution{C}{}\LateralR{\mathcal{C}}}}{X\in\Dom{\substitution{C}{}}}.
            \]
        \end{definition}
    
        A close inspection of the previous definition and the definition of lateral substitutions ensures that the set just introduced is a valid set of AUTs.\
        Suppose \(\mathcal{C}\OneStep{}\mathcal{C}'\).\
        If \SOLNRname\ is the employed rule, then \(\ext{C}{}'=\ext{C}{}\).\
        If any other rule is the employed one, then \(\ext{C}{}'=\ext{C}{}\cup\Set{\eAUT{X}}\), where \(\eAUT{X}\) is the AUT in \(\unsolved{C}{}\) that is transformed by the rule.\
        Thus, an inductive argument together with Lemma \ref{lemma::Problem-Invariance} provides the following natural result: 
    
        \begin{lemma}[Preservation of Extended Set]\label{lemma::Preservation-of-Extended-Set}
            If \(\mathcal{C}\derives\mathcal{C}'\), then \(\ext{C}{}\subseteq\ext{C}{}'\).
        \end{lemma}
    
        Notice that in the set of labels of a valid configuration, the labels of the unsolved and solved parts are obtained directly from the AUTs that are already present in the configuration.\
        The labels in the domain of the substitution refer to AUTs that were already transformed, and the extended set of AUTs is a very natural way to recover them.\
        Thus, one can define:
    
        \begin{definition}[Total Set of AUTs]
            The \emph{total set of AUTs} of a valid configuration \(\mathcal{C}\) is defined by
            \[
                \tot{C}{}\coloneqq\unsolved{C}{}\cup\solved{C}{}\cup\ext{C}{}.
            \]
        \end{definition}
    
        A close inspection of the previous definition ensures that the set just introduced is a valid set of AUTs.\
        Moreover, the identity \(\labs{\tot{C}{}}=\labs{\mathcal{C}}\) holds.\
        Suppose \(\mathcal{C}\OneStep{}\mathcal{C}'\).\
        If \SOLNRname, \SOLRname, or \SYNTname\ is the employed rule, then \(\tot{C}{}'=\tot{C}{}\).\
        If any decomposition rule is the employed one, then \(\ext{C}{}'=\ext{C}{}\cup A\), where \(A\) is the set of new AUTs introduced by the employed rule.\
        Thus, an inductive argument provides the following natural result:
    
        \begin{lemma}[Preservation of Total Set]\label{lemma::Preservation-of-Total-Set}
            If \(\mathcal{C}\derives\mathcal{C}'\), then \(\tot{C}{}\subseteq\tot{C}{}'\).
        \end{lemma}

        Given two terms \(s\) and \(t\), a term \(g\) is a generalizer for \(s\) and \(t\) iff there exist substitutions \(\sigma\) and \(\tau\) such that \(g\sigma=s\) and \(g\tau=t\).\
        Thus, a generalizer for an AUT is a generalizer for both its sides.\
        In the case of a valid configuration, one requires a notion of a generalizer that \emph{uniformly generalizes} all the AUTs associated with that configuration.\
        Thus, a generalizer for a valid configuration must be a substitution, and any comparison between generalizers must rely on the instantiation preorder restricted to the set of labels of the configuration.\
        A natural definition is the following one:

        \begin{definition}[Total Generalizer for a Valid Configuration]\
            \begin{enumerate}[label=\textbf{\arabic*.}]
                
                \item Let \(A\) be a valid set of AUTs.\
                A substitution \(\gamma\) is a \emph{total generalizer for} \(A\) iff there exist \emph{left} and \emph{right cohesion substitutions} \(\GLateralL{A}\) and \(\GLateralR{A}\), respectively, such that
                \begin{align*}
                    X\gamma\GLateralL{A}&=\lhs{\eAUT{X}}\quad\quad\text{and}\quad\quad X\gamma\GLateralR{A}=\rhs{\eAUT{X}}
                \end{align*}
                for every \(\eAUT{X}\) in \(A\);

                \item A substitution \(\gamma\) is a \emph{total generalizer for} a valid configuration \(\mathcal{C}\) iff \(\gamma\) is a total generalizer for \(\tot{C}{}\).\
                Its left and right cohesion substitutions are denoted by \(\GLateralL{\mathcal{C}}\) and \(\GLateralR{\mathcal{C}}\), respectively.
                
            \end{enumerate}
        \end{definition}

        That is the notion of a generalizer for a valid configuration required for proving completeness.\
        The identity \(\iota\) is a total generalizer for every valid configuration \(\mathcal{C}\) and the lateral substitutions of \(\tot{C}{}\) are its cohesion substitutions.\
        That mirrors the fact that a variable generalizes any term.\
        As a bonus, the new notions presented so far are already enough to prove soundness in a way that is different from the ones presented in \cite{DBLP:conf/nfm/AyalaRinconLLMK25} or in \cite{DBLP:journals/iandc/AlpuenteEEM14}.\
        The new proof of soundness relies on the preservation properties stated in Lemmas \ref{lemma::Problem-Invariance} and \ref{lemma::Preservation-of-Total-Set}, and it is done by induction on the number of steps of the derivation.
        
        The pen-and-paper proof of completeness that is going to be presented here is constructive and completely different from the one presented in \cite{DBLP:journals/iandc/AlpuenteEEM14}, because no results on confluence or the type of syntactic anti-unification will be assumed.\
        On the contrary, that type will be a consequence of termination, soundness, and completeness.\
        Before presenting the proof, certain properties of generalizers for valid configurations must be proved.\
        The first one is a natural definitional~corollary:

        \begin{lemma}[Total Generalizer Coherence]\label{lemma::Generalizer-Coherence}
            If \(A\) is a valid set of AUTs, \(\gamma\) is a total generalizer for it, and \(\eAUT{X}\) is a member of \(A\), then \(X\gamma\) is a total generalizer for \(\eAUT{X}\).\
            More precisely:
            \begin{enumerate}[label=\textbf{\arabic*.}]

                \item If \(\eAUT{X}\) is a decomposable AUT, then:
                \begin{enumerate}[label=\textbf{1.\arabic*.}]
                
                    \item If \(\lhs{\eAUT{X}}\) is a functional application, then \(X\gamma\) is either a variable or a functional application starting with the same root symbol of \(\lhs{\eAUT{X}}\);

                    \item If \(\lhs{\eAUT{X}}\) is a pair, then \(X\gamma\) is either a variable or a pair;
                    
                \end{enumerate}

                \item If \(\eAUT{X}\) is a trivial AUT, then \(X\gamma\) is the unit, a constant or a variable;

                \item If \(\eAUT{X}\) is a solved AUT, then \(X\gamma\) is a variable;
                
            \end{enumerate}
        \end{lemma}

        A particularly important preservation result concerning total generalizers for valid configurations is the following one.\
        Despite being a natural definitional corollary, it essentially informs that a total generalizer cannot associate the same solution to problems encoded by AUTs of distinct classes:

        \begin{lemma}[Preservation of Classification]\label{lemma::Preservation-Classification}
            Let \(A\) be a valid set of AUTs and \(\gamma\) be an arbitrary total generalizer for \(A\).\
            For all \(\eAUT{X}\) and \(\eAUT{Y}\) in \(A\), if \(X\gamma=Y\gamma\), then \(\eAUT{X}\) and \(\eAUT{Y}\) are repeated.
        \end{lemma}

        The key step in the proof of the previous lemma is the existence of cohesion substitutions associated with the total generalizer \(\gamma\).\
        Under the hypothesis \(X\gamma=Y\gamma\), they force the left and right-hand sides of the AUTs \(\eAUT{X}\) and \(\eAUT{Y}\) to be respectively the same terms.\
        A particularly important consequence of the previous lemma is the following lemma, which deals with syntactic and solved AUTs:

        \begin{lemma}[Separation]\label{lemma::Separation}
            Let \(A\) be a valid set of AUTs, \(\gamma\) be an arbitrary total generalizer for \(A\), and \(\eAUT{X}\) and \(\eAUT{Y}\) be arbitrary members of \(A\).
            \begin{enumerate}[label=\textbf{\arabic*.}]

                \item If \(\eAUT{X}\) and \(\eAUT{Y}\) are trivial AUTs and \(\lhs{\eAUT{X}}\neq\lhs{\eAUT{Y}}\), then \(X\gamma\neq Y\gamma\);

                \item If all members of \(A\) are not repeated in \(A\) and \(X\neq Y\), then \(X\gamma\neq Y\gamma\).

            \end{enumerate}

        \end{lemma}

        An important application of Lemmas \ref{lemma::Generalizer-Coherence} and \ref{lemma::Separation} is the following.\ If \(\mathcal{C}\) is a valid configuration, then the members of \(\solved{C}{}\) are not repeated in it.\
        Thus, a total generalizer for \(\mathcal{C}\) maps distinct labels in \(\labs{\solved{C}{}}\) to distinct variables.\
        In other words, a total generalizer for a valid configuration is able to separate non-repeated solved (resp.\ non-repeated trivial) AUTs that appear in a derivation ending in that same configuration.\
        This property will play an important role in a step of the proof of completeness.\
        Another important property of total generalizers for valid configurations is presented in the following lemma:

        \begin{lemma}[Choice of Total Generalizer]\label{lemma::Choice-Generalizer}
             Let \(\mathcal{C}\) be a valid configuration and \(\mathbb{A}\) be a set of variables.\
             For every total generalizer \(\gamma\) for \(\mathcal{C}\) there exists a total generalizer \(\gamma'\) for \(\mathcal{C}\) such that \(\RVar{\gamma'}\cap\mathbb{A}=\emptyset\) and \(\gamma'\simeq_{\labs{\mathcal{C}}}\gamma\).
        \end{lemma}

        To prove the previous lemma, one needs to enumerate all the variables in \(\RVar{\gamma}\cap\mathbb{A}\) (if there are any) and rename all of them into distinct fresh variables using a renaming \(\alpha\).\
        New cohesion substitutions are constructed using \(\alpha\) and the old ones.\
        Lemma \ref{lemma::Choice-Generalizer} serves as inspiration for providing the following definition:

        \begin{definition}[Restricted Total Generalizer]\label{definition:restrictedtotalgeneralizer}
            Let \(\mathcal{C}\) be a valid configuration and \(\mathcal{C}_f\) be a final configuration such that \(\mathcal{C}\derives\mathcal{C}_f\).\
            A total generalizer \(\gamma\) for \(\mathcal{C}\) is a \emph{restricted total generalizer for} \(\mathcal{C}\) \emph{relative to} \(\mathcal{C}_f\) iff \(\RVar{\gamma}\cap\left(\labs{\mathcal{C}_f}\cup\Protected{\mathcal{C}_f}\right)=\emptyset\) and \(\gamma\gamma=\gamma\).
        \end{definition}

        The specification of that type of total generalizer is crucial in the proof of completeness.\
        Lemma \ref{lemma::Choice-Generalizer} shows that if a valid configuration admits a total generalizer, then it admits a restricted total generalizer that is equivalent to the original one on the labels of that configuration.\
        That means that one can restrict their attention to that class of more well-behaved total generalizers for valid configurations.\
        Also, since equivalence holds, completeness is not lost by such a restriction.\
        That is the path followed in the PVS specification.
        We are ready to state completeness.\
        All the preservation properties, invariants, and (restricted) total generalizer properties presented so far will naturally guide the construction of the proof.\
        It will be dismembered into parts.\
        \SYNTname, \SOLNRname, and \SOLRname\ rules are presented in Subsection~\ref{subsection:Completeness-Analysis-for-Solve-Syn-Rules}, while \DECFname\ and \DECPname\ are presented in Subsection~\ref{subsection:Completeness-Analysis-for-Decomposition-Rules}.

        \begin{theorem}[Strong Completeness for Restricted Total Generalizers]\label{theorem::Strong-Completeness-Restricted-Generalizers}
            Let \(\mathcal{C}_0\) be an initial configuration, \(\mathcal{C}_f\) be a final configuration such that \(\mathcal{C}_0\derives\mathcal{C}_f\), and \(\theta_f\) be the computed substitution of \(\mathcal{C}_f\).\
            For every valid configuration \(\mathcal{C}\) such that \(\mathcal{C}_0\derives\mathcal{C}\derives\mathcal{C}_f\) and every restricted total generalizer \(\gamma\) for \(\mathcal{C}\) relative to \(\mathcal{C}_f\) it is the case that \(\gamma\preceq_{\labs{\mathcal{C}}}\theta_f\).
        \end{theorem}

        The proof of Theorem~\ref{theorem::Strong-Completeness-Restricted-Generalizers} is done by induction on the number of steps taken in \(\mathcal{C}\derives\mathcal{C}_f\).\
        In the base case, one has \(\mathcal{C}=\mathcal{C}_f\).\
        Let \(S_f\) be the solved part of \(\mathcal{C}_f\).\
        By the definition of total generalizer for \(\mathcal{C}_f\), one can always assume that \(\Dom{\gamma}\cap\Protected{\mathcal{C}_f}=\emptyset\).\
        Since \(\RVar{\theta_f}\subseteq\labs{S_f}\cup\Protected{\mathcal{C}_f}\), one employs the definition of restricted total generalizer and Lemmas~\ref{lemma::History-Solved}, \ref{lemma::History-Domain}, \ref{lemma::Generalizer-Coherence}, and \ref{lemma::Separation} to conclude that~\(\gamma\preceq_{\labs{\mathcal{C}_f}}\theta_f\).
        
        \begin{figure} [htpb]
            \centering
            \begin{tikzpicture}[scale=1]
                \draw (0,0) node {\(\boldsymbol{\mathcal{C}_0}\)};
                \draw (4,0) node {\(\boldsymbol{\color{purple}\mathcal{C}}\)};
                \draw (8,0) node {\(\boldsymbol{\color{blue}\mathcal{C}'}\)};
                \draw (12,0) node {\(\boldsymbol{\color{blue!30!green!100!black!100!}\mathcal{C}_f}\)};
                \draw[thick,double,-myarrow] (0.3,0) -- (3.7,0) node[midway, above] {\(\boldsymbol\ast\)};
                \draw[thick,double,-myarrow] (4.25,0) -- (7.65,0) node[midway, above] {\scriptsize{after one rule step}};
                \draw[thick,double,-myarrow] (8.25,0) -- (11.65,0) node[midway, above] {\scriptsize{\(\boldsymbol m\) steps}};
                \draw [color=purple,thick,decorate,decoration={brace,amplitude=5pt,raise=4ex}]
      (0,0) -- (4,0) node[midway,yshift=3em]{{\scriptsize\color{black}given} \(\color{purple}\boldsymbol\gamma\)};
                \draw [color=blue,thick,decorate,decoration={brace,amplitude=5pt,mirror,raise=4ex}]
      (0,0) -- (8,0) node[midway,yshift=-3.5em]{{\scriptsize\color{black}from} \(\boldsymbol{\color{purple}\gamma}\) {\scriptsize\color{black}construct} \(\boldsymbol{\color{blue}\gamma'}\)};
                \draw [color=blue!30!green!100!black!100!,thick,decorate,decoration={brace,amplitude=5pt,mirror,raise=10ex}]
      (0,0) -- (12,0) node[midway,yshift=-6em]{{\scriptsize\color{black}from} \(\boldsymbol{\color{blue}\gamma'\color{black}\preceq_{\labs{{\color{blue}\mathcal{C}'}}}\color{blue!30!green!100!black!100!}\theta_f}\) {\scriptsize\color{black}prove} \(\boldsymbol{\color{purple}\gamma\color{black}\preceq_{\labs{{\color{purple}\mathcal{C}}}}\color{blue!30!green!100!black!100!}\theta_f}\)};
            \end{tikzpicture}
            \caption{Inductive step in the proof of completeness. } \label{figure::Completeness}
        \end{figure}
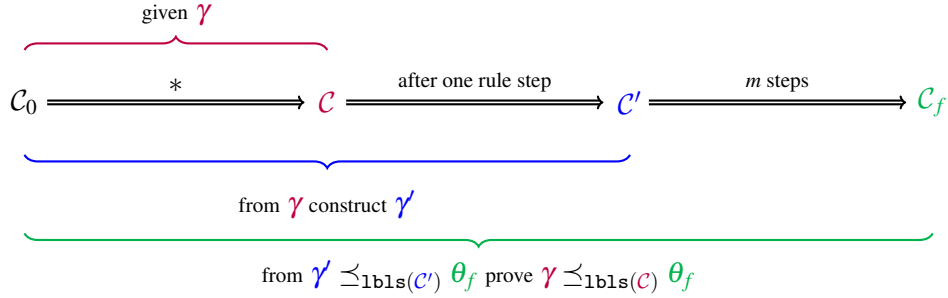
        
        For the inductive step, one assumes the result for derivations of length \(m\) and supposes that \(\mathcal{C}_0\derives\mathcal{C}\Rightarrow^{m+1}\mathcal{C}_f\).\
        Thus, there exists a valid configuration \(\mathcal{C}'\) such that \(\mathcal{C}_0\derives\mathcal{C}\OneStep{}\mathcal{C}'\Rightarrow^{m}\mathcal{C}_f\).\
        The analysis now proceeds by considering the rule employed in the step \(\mathcal{C}\OneStep{}\mathcal{C}'\).\
        One must start with an arbitrary restricted total generalizer \(\gamma\) for \(\mathcal{C}\) relative to \(\mathcal{C}_f\) and use it to construct a restricted total generalizer \(\gamma'\) for \(\mathcal{C}'\) relative to \(\mathcal{C}_f\) to use the induction hypothesis (i.h.).\
        That construction must be so that one can push the comparison using the instantiation preorder back to \(\gamma\).\
        The situation is presented in Figure \ref{figure::Completeness}.\
        Once the proof is finished, one can employ Lemma \ref{lemma::Choice-Generalizer} to derive the following corollary from which one concludes that the type of syntactic anti-unification is unitary:

        \begin{corollary}[Strong Completeness]\label{corollary::Strong-Completeness}
            Let \(\mathcal{C}_0\) be an initial configuration, \(\mathcal{C}_f\) be a final configuration such that \(\mathcal{C}_0\derives\mathcal{C}_f\), and \(\theta_f\) be the computed substitution of \(\mathcal{C}_f\).\
            For every valid configuration \(\mathcal{C}\) such that \(\mathcal{C}_0\derives\mathcal{C}\derives\mathcal{C}_f\) and every total generalizer \(\gamma\) for \(\mathcal{C}\) it is the case that \(\gamma\preceq_{\labs{\mathcal{C}}}\theta_f\).
        \end{corollary}

\subsection{Completeness Analysis for Syntactic and Solve Rules}\label{subsection:Completeness-Analysis-for-Solve-Syn-Rules}

        This subsection is concerned with the branches of the proof of completeness where the step \(\mathcal{C}\OneStep{}\mathcal{C}'\) is achieved by means of the rules \SYNTname, \SOLNRname, or \SOLRname.\
        Let \(\gamma\) be an arbitrary restricted total generalizer for \(\mathcal{C}\) relative to \(\mathcal{C}_f\).\
        The information concerning both \(\gamma\) and the history of the computation is required for constructing a restricted total generalizer \(\gamma'\) for \(\mathcal{C}'\) relative to \(\mathcal{C}_f\) that allows the use of the reasoning illustrated in Figure \ref{figure::Completeness}.
        \begin{enumerate}[label=\textbf{\arabic*.}]

            \item If \SYNTname\ is the employed rule, then there exists a trivial AUT \(\eAUT{X}\) in \(\unsolved{C}{}\) such that \(\unsolved{C}{}'=\unsolved{C}{}\setminus\Set{\eAUT{X}}\), \(\solved{C}{}'=\solved{C}{}\), and \(\substitution{C}{}'=\substitution{C}{}\BasicSub{X}{\lhs{\eAUT{X}}}\).\
            By Lemma \ref{lemma::Preservation-Classification}, one knows that \(X\gamma\) is different from every variable that is assigned by \(\gamma\) to the labels of non-trivial AUTs.\
            The problem encoded by the trivial AUT \(\eAUT{X}\) may appear in different positions in the common term structure. The information about it may be already stored in \(\unsolved{C}{}\setminus\Set{\eAUT{X}}\) or in~\(\substitution{C}{}\).
            
            \item If \SOLNRname\ is the employed rule, then there exists a solved AUT \(\eAUT{X}\) in \(\unsolved{C}{}\) such that \(\unsolved{C}{}'=\unsolved{C}{}\setminus\Set{\eAUT{X}}\), \(\solved{C}{}'=\solved{C}{}\cup\Set{\eAUT{X}}\), and \(\substitution{C}{}'=\substitution{C}{}\).\
            In other words, this is the first time that the problem encoded by the solved AUT \(\eAUT{X}\) is encountered.\
            From Lemma \ref{lemma::Separation}, it follows that \(X\gamma\) is different from \(Y\gamma\) for every \(Y\) in \(\labs{\solved{C}{}}\).\
            By Lemma \ref{lemma::Preservation-Classification}, one knows that \(X\gamma\) is different from every variable that is assigned by \(\gamma\) to the labels of non-solved AUTs.\
            Notice that no information about the problem is already stored in~\(\substitution{C}{}\).\
            Nonetheless, information about it may already be stored in~\(\unsolved{C}{}\setminus\Set{\eAUT{X}}\).

            \item If \SOLRname\ is the employed rule, then there exist solved AUTs \(\eAUT{X}\) in \(\unsolved{C}{}\) and \(\eAUT{X^{\prime}}\) in \(\solved{C}{}\) such that \(\unsolved{C}{}'=\unsolved{C}{}\setminus\Set{\eAUT{X}}\), \(\solved{C}{}'=\solved{C}{}\), and \(\substitution{C}{}'=\substitution{C}{}\BasicSub{X}{X^{\prime}}\).\
            In other words, the problem encoded by the solved AUT \(\eAUT{X^{\prime}}\) is encountered again.\
            From Lemma \ref{lemma::Separation}, it follows that \(X\gamma\) and \(X^{\prime}\gamma\) are different from \(Y\gamma\) for every \(Y\) in \(\labs{\solved{C}{}}\setminus\Set{X^{\prime}}\).\
            Since \(\eAUT{X}\) and \(\eAUT{X^{\prime}}\) are repeated AUTs, it is not necessarily the case that \(X\gamma=X^{\prime}\gamma\).\
            By Lemma \ref{lemma::Preservation-Classification}, one knows that \(X\gamma\) is different from every variable that is assigned by \(\gamma\) to the labels of non-solved AUTs.\
            The problem encoded by the solved AUT \(\eAUT{X}\) may appear in different positions in the common term structure, and the information about it is already stored in~\(\solved{C}{}\).
            
        \end{enumerate}

        That explains why the elements provided by \cite{DBLP:conf/nfm/AyalaRinconLLMK25} are not sufficient for addressing those rules and why it is crucial to require in the specification of a total generalizer for a valid configuration that it must also generalize both its solved part and the domain of its computed substitution together with its unsolved part.\
        Only requiring that for the latter does not necessarily provide all the needed information regarding the history of the computation to compare the restricted total generalizer and the final computed substitution in those branches of the proof.\
        Now, both the definition of total generalizer and the preservation properties previously discussed come to the rescue.\
        Since the employed rule is \SYNTname, \SOLNRname, or \SOLRname, one knows by Lemma \ref{lemma::Preservation-of-Total-Set} that \(\tot{C}{}'=\tot{C}{}\) and \(\labs{\mathcal{C}'}=\labs{\mathcal{C}}\).\
        In other words, all the information required for the comparison is already present in \(\mathcal{C}\) and is preserved from \(\mathcal{C}\) to \(\mathcal{C}'\).\
        Thus, one concludes that \(\gamma\) is already a restricted total generalizer for \(\mathcal{C}'\) relative to \(\mathcal{C}_f\) and \(\gamma\preceq_{\labs{\mathcal{C}}}\theta_f\) follows immediately from the induction hypothesis.\
        Although the pen-and-paper proof seems straightforward, much of the effort in the verification will be related to its dependencies.
        
\subsection{Completeness Analysis for Decomposition Rules}\label{subsection:Completeness-Analysis-for-Decomposition-Rules}

        This subsection is concerned with the branches of the proof of completeness where the step \(\mathcal{C}\OneStep{}\mathcal{C}'\) is achieved by means of either \DECFname\ or \DECPname.\
        Let \(\gamma\) be an arbitrary restricted total generalizer for \(\mathcal{C}\) relative to \(\mathcal{C}_f\).
        \begin{enumerate}[label=\textbf{\arabic*.}]

            \item If \DECFname\ is the employed rule, then there exist a decomposable AUT \(\eAUT{X}=\auEq{X}{ft_L}{ft_R}\) in \(\unsolved{C}{}\) and a variable \(Y\) fresh for \(\mathcal{C}\) such that \(\unsolved{C}{}'=\unsolved{C}{}\cup\Set{\eAUT{Y}}\setminus\Set{\eAUT{X}}\), \(\solved{C}{}'=\solved{C}{}\), \(\substitution{C}{}'=\substitution{C}{}\BasicSub{X}{fY}\), and \(\eAUT{Y}=\auEq{Y}{t_L}{t_R}\).\
            By Lemma \ref{lemma::Generalizer-Coherence}, the term \(X\gamma\) is either a variable or \(ft\) for some term \(t\).

            \begin{enumerate}[label=\textbf{1.\arabic*.}]

                \item Suppose \(X\gamma=ft\) for some term \(t\).\ Define:
                \[
                    \gamma'\coloneqq\res{\gamma}{\Dom{\gamma}\setminus\Set{Y}}\BasicSub{Y}{t}.
                \]
                That is a restricted total generalizer for \(\mathcal{C}'\) relative to \(\mathcal{C}_f\).\
                By the i.h., \(\gamma'\preceq_{\labs{\mathcal{C}'}}\theta_f\).\
                There exists a substitution \(\delta'\) such that \(\gamma'\delta'=_{\labs{\mathcal{C}'}}\theta_f\).\
                Letting
                \[
                    \delta\coloneqq\BasicSub{Y}{t}\delta',
                \]
                one concludes that \(\gamma\delta=_{\labs{\mathcal{C}}}\theta_f\).\
                Indeed,
                \begin{align*}
                    X\gamma\delta&=ft\delta'=fY\gamma'\delta'=fY\theta_f=X\theta_f,
                \end{align*}
                and \(Z\gamma\delta=Z\gamma'\delta'\) for \(Z\) in \(\labs{\mathcal{C}}\setminus\Set{X}\).\
                Thus, \(\gamma\preceq_{\labs{\mathcal{C}}}\theta_f\);

                \item Suppose \(X\gamma=U\) is a variable.\
                There exist cohesion substitutions \(\GLateralL{\mathcal{C}}\) and \(\GLateralR{\mathcal{C}}\) of \(\gamma\).\
                There exists a substitution \(\delta''\) such that \(\gamma\delta''=_{\labs{\mathcal{C}}}\theta_f\gamma\).\
                Let \(W\) be fresh for \(\mathcal{C}_f\), \(\gamma\), \(\GLateralL{\mathcal{C}}\), \(\GLateralR{\mathcal{C}}\), and \(\delta'\).\
                Define:
                \[
                    \gamma'\coloneqq\res{\gamma}{\Dom{\gamma}\setminus\Set{Y}}\BasicSub{U}{fW}\BasicSub{Y}{W}.
                \]
                That is a restricted total generalizer for \(\mathcal{C}'\) relative to \(\mathcal{C}_f\).\
                By the i.h., \(\gamma'\preceq_{\labs{\mathcal{C}'}}\theta_f\).\
                There exists a substitution \(\delta'\) such that \(\gamma'\delta'=_{\labs{\mathcal{C}'}}\theta_f\).\
                Letting
                \[
                    \delta\coloneqq\BasicSub{U}{fW}\BasicSub{Y}{W}\delta',
                \]
                one concludes that \(\gamma\delta=_{\labs{\mathcal{C}}}\theta_f\).\
                Indeed,
                \begin{align*}
                    X\gamma\delta&=fW\delta'=fY\gamma'\delta'=fY\theta_f=X\theta_f,
                \end{align*}
                and \(Z\gamma\delta=Z\gamma'\delta'\) for \(Z\) in \(\labs{\mathcal{C}}\setminus\Set{X}\).\
                Thus, \(\gamma\preceq_{\labs{\mathcal{C}}}\theta_f\);

            \end{enumerate}

            \item The analysis for \DECPname\ is similar and will be omitted.
            
        \end{enumerate}

        Crucial steps in verifying the previous arguments rely on the constraints imposed by the definition of restricted total generalizers.\
        For instance, the constraint \(\RVar{\gamma}\cap\left(\labs{\mathcal{C}_f}\cup\Protected{\mathcal{C}_f}\right)=\emptyset\) is vital for the construction of the substitution \(\gamma'\).\
        Moreover, the fact that a total generalizer for a valid configuration generalizes the unsolved part of that configuration is crucial for applying the induction step in verifying these branches of the proof.

\section{Conclusion}\label{sec:conclusion}

This paper presented a rigorous pen-and-paper proof of the completeness theorem for syntactic anti-unification and the discussion clarified the formal elements required to mechanically verify the theorem inductively in PVS (and other proof assistants as well). In particular, the paper highlighted important differences that make that mechanization exercise much more complex than verifying the algorithmic solutions to unification, for which the cases of solved and syntactically equal subproblems are resolved immediately, leading to failures and trivial solutions.
For formalizing the proof of the completeness theorem in PVS, we addressed these problems with the necessary theoretical rigor. The greater complexity of mechanizing verification for anti-unification algorithms lies in how to record the history of decomposing problems into solved and syntactic subproblems, whose regularities (repetitions) are further encoded by storing repeated solved and syntactic subproblems via the substitution solution under construction. The precise notions that guarantee the construction of a solution more specific than any arbitrary generalizer involve elaborate invariant properties related to the preservation of the problem (Lemma \ref{lemma::Problem-Invariance}),  history of the computed solution (Lemma \ref{lemma::History-Computed-Substitution}),  preservation of the extended problem and coherence of the partial solution generated during the computation (Lemmas \ref{lemma::Preservation-of-Total-Set}, and \ref{lemma::Generalizer-Coherence}).  In addition, by definition of a restricted notion of arbitrary generalizers (Definition \ref{definition:restrictedtotalgeneralizer}), the main completeness result is obtained as a corollary (Corollary \ref{corollary::Strong-Completeness}) of the completeness for such a class of generalizers (Theorem \ref{theorem::Strong-Completeness-Restricted-Generalizers}).
All results presented in Subsection \ref{subsection:newrequiredDefinitions} are already completely specified and formally verified in PVS.

\subsection*{Future work} Work in progress includes the formalization in PVS of the definitions and lemmas presented in Section~\ref{subsection:Adaptation-of-Generalizer-Notions} to obtain a mechanically and formally verified proof of completeness for the functional rule-based algorithm \texttt{Antiunify}, from which certified executable code can be extracted.\
Additional work includes the extension of the formalization for verifying algorithms to anti-unification modulo various algebraic properties widely used in mathematics, such as commutativity, associativity \cite{DBLP:conf/jelia/AlpuenteEEM14,DBLP:journals/amai/AlpuenteEMS22}, absorption \cite{DBLP:conf/ijcar/AyalaRinconCBK24,AFGonzalezBarragan2025}, and their combinations.

\subsubsection*{Acknowledgments} This work was partially supported by the Austrian Science Fund (FWF) project P 35530, the Rustaveli National Science Foundation of Georgia under project FR-25-5957, the Brazilian Research Council CNPq Universal grant 407461/2025-6, and Research productivity grants 313290/2021-0 and 6/2026-7, and a PDSE scholarship of the Brazilian Higher Education Council (CAPES) Finance Code 001 to the last author.

\bibliographystyle{plain}

\end{document}